\documentstyle[12pt,aaspp4]{article}

\received{}
\accepted{}
\slugcomment{Submitted to Nature}

\begin{document}
\include{psfig}

\def\kms{km~s~$^{-1}$ }
\def\Lya{Ly$\alpha$ }
\def\lya{Ly$\alpha$ }
\def\Lyb{Ly$\beta$ }
\def\lyb{Ly$\beta$ }
\def\Lyg{Ly$\gamma$ }
\def\Lyd{Ly$\delta$ }
\def\Lye{Ly$\epsilon$ }
\def\Ly{Lyman~}
\def\ang{\AA }
\def\gq{$\geq$ }
\def\zem{$z_{em}$ }
\def\zabs{$z_{abs}$ }

%
\title{The Cosmological Baryon Density from the Deuterium Abundance 
at a redshift $z = 3.57$}

\author{DAVID TYTLER\altaffilmark{1}, XIAO-MING FAN\altaffilmark{1} 
	\& SCOTT BURLES\altaffilmark{1}}
\affil{Department of Physics, and Center for Astrophysics and Space
Sciences \\
University of California, San
Diego \\
C0111, La Jolla, CA 92093-0111}
 
\altaffiltext{1}{Visiting Astronomer, W. M. Keck Telescope, California
Association for Research in Astronomy}
 
%
%
 

The primordial ratio of the 
number of deuterium to hydrogen nuclei (D/H) created in 
big bang nucleosynthesis is the
most sensitive measure of the cosmological baryon to 
photon ratio and the
cosmological density of baryons $\Omega_b$ (\cite{wal91}-\cite{cop95}).
In the interstellar medium (ISM) of our Galaxy 
D/H $= 1.6 \pm 0.1 \times 10^{-5}$ (\cite{lin95}), 
which places a strict lower limit on the primordial abundance,
because stars reduce the proportion of D in the ISM.
Quasar absorption systems (QAS) should give definitive measurements of
the primordial D because they can sample metal-poor gas at early epochs 
where the destruction of D should be negligible.  
A probable measurement of a high abundance ratio was reported in one QAS: 
D/H $\approx 24 \times 10^{-5}$ (\cite{son94},\cite{car94}).
Here we report a measurement of low D/H = 
2.3 $\times 10^{-5}$, in another QAS,
which is consistent with D/H in the ISM, with models of 
Galactic chemical evolution, 
(\cite{edm94},\cite{ste95}), 
and with earlier data 
(\cite{son94},\cite{car94}) provided that 
D line is strongly contaminated.
We have internal consistency checks, and believe that
this is the first direct measurement of a primordial abundance ratio. 
It provides the most accurate measurement of the baryon density: a high
value of $\Omega_b = 0.05$, assuming a Hubble constant 
$H_0$ = 70 km s$^{-1}$ Mpc$^{-1}$.

Deuterium is rarely seen in QAS because the associated hydrogen 
is usually distributed over 
a wide range of velocities and its absorption covers deuterium. 
We selected QSO 1937--1009 because we found that
the absorption system at $z=3.572$ showed
unusually weak metal lines in low resolution spectra,
which we took to indicate that the gas was
distributed in one or two narrow velocity components (\cite{pet90}).
We obtained optical spectra of QSO 1937--1009 ($z_{em} = 3.78$, 
V $\simeq 17.5$) with the HIRES echelle spectrograph 
(\cite{vog94}) on the W. M. Keck 10-m telescope.
The high quality spectra make this amongst the best characterized QAS, and
show strong absorption at the expected position of the D 
\Lya line and weak but highly significant absorption at D \Lyb.
 
The weak metal lines in the system, 
C~II, C~IV, Si~II, and Si~IV, appear to
have the same profiles, which are adequately described by two components
separated by 15 km~s$^{-1}$. 
The velocity positions determined from the metals (Figure 1)
lie in the cores of
the higher order Lyman lines (Figure 2), and at the expected positions of the
deuterium absorption features.
The absorption features are modeled assuming that 
the velocities of these two components, labeled blue (zero velocity) and red,
are common to every ion.
The parameters and their statistical errors which produce the
smooth curves in Figures 1 \& 2  are shown in Table 1. 

The Lyman lines show that $>90$\% of the H~I must be in, near, or
between the two components, but not elsewhere because there are no other
lines in the spectrum which can explain the Lyman limit.
A simultaneous fit to the two components
gives a total column density Log N(H~I) = 17.94 $\pm$ 0.05. 
D/H will change if the total H changes, but not if we redistribute the H
in the velocity range where the D is measured, because D is unsaturated.  
There could be additional H 
at velocities between, or near these components, which would need $b< 20$ 
($b \equiv \sqrt{2} \sigma$ is a measure of the intrinsic, 
not instrumental, line width).
But this can not significantly increase the total N(H~I), because the wings 
of \Lya require Log N(H~I) $<$ 18.04.
We can not decrease the total N(H~I) because we must completely absorb the 
cores of the high order Lyman lines, and in the Lyman continuum.
An additional component at $v \simeq +49$ \kms
produces S~IV, C~IV and H~I absorption, but it
is absent from the high order Lyman lines,
it has a low column density of Log N(H~I) = 14.94 $\pm$ 0.14, and
it can not change D/H.

The red D component is blended with both the blue D and H components, hence
its parameters are poorly constrained by the data.
We fit both components simultaneously and require that
(D/H)$_{red}$=(D/H)$_{blue}$=(D/H)$_{total}$.
We also restrict its $b$ value to the range allowed by $b$(H).
The total D in both components is Log $N$(D~I) $= 13.30 \pm 0.04$, which is 
fairly insensitive to $b$(D) and the precise velocity of the D
because the D lines are unsaturated.
We then have Log D/H $= -4.64 \pm 0.06$, where this error includes
random photon noise, and fitting errors, but not systematic errors.

Statistically we expect some H in the region of the D lines, because
weak H~I absorption is extremely common. This will 
systematically increase N(D).
We generated noise free model spectra  for various N(D),
using the redshifts, $b$-values, and the N(H~I) determined above.
To each spectrum we then added \Lya forest lines
with random N, $b$ and $z$ values drawn from known distributions  
(\cite{kir96},\cite{hu95}).
For a given D/H, we made 500,000 random spectra, and
we calculated the likelihood that the data 
came from the realizations of that model.
The maximum likelihood occurred for the measured D/H with no
significant \Lya contamination.
The likelihood function has a long tail to lower D/H with an
expected (mean) value of 
D/H $= 2.27 ^{+0.09}_{-0.14} \times 10^{-5}$, 
($1\sigma $ errors from H blended with N(D), and photon errors at D),
only slightly lower than the D/H of maximum likelihood. 
We conclude that in this QAS, random \Lya lines are unlikely to give
profiles that look like the observed D, and that the absorption we 
measure is entirely deuterium to within the random errors.

We estimated the remaining 
systematic errors from the uncertainty in the continuum 
level by moving the continuum $\pm$ 2\% 
at the Ly-$\alpha$ feature, and $\pm$ 5\% at the Lyman Limit.
These percentage adjustments correspond to the $1\sigma $ error per pixel
in each region of the spectrum. The random error in the
continuum level should be much less than this, because the continuum is
constant over many pixels.
The same fitting procedure was run with each new continuum level.
When the continuum was raised in both regions we found
$\Delta \rm log \it N \rm(H~I) = {+0.04}$
and $\Delta$ log N(D~I) = $-0.02$, and when it was lowered,
$\Delta \rm log \it N(\rm H~I) = {-0.03}$ and 
$\Delta \rm log \it N(\rm D~I) = +0.02$. 
We conclude that 
\begin{displaymath}
\rm Log \, (D/H) = -4.64 \pm 0.06 \; ^{+0.05}_{-0.06}, 
\end{displaymath}
\begin{displaymath}
or ~~~~~~~
D/H = 2.3 \, \pm 0.3 \, \pm 0.3 \times 10^{-5},
\end{displaymath}
where the log form is more accurate and
the errors represent the 1$\sigma$ statistical uncertainty, followed by
the systematic uncertainty.

 
As a consistency check, we used the measured $b$ values of the C~II, Si~II 
and hydrogen lines
to find the temperature and turbulent velocity of each component.
We assume that all lines of these ions have the same turbulent widths
$b_{tur}$, and temperature. 
The intrinsic line widths are $b_{total} = (b_{therm}^2 + b_{tur}^2)^{0.5}$,
where the thermal widths depend on the mass of each ion:
$b_{therm} = 0.128 (T m_p/m_{ion})^{0.5}$ km s$^{-1}$, 
where $m_p$ is the proton mass,
and values are given in the Table.
We then predict that the blue component of D has
$b = 12.5 \pm 2.1$ km s$^{-1}$, which is consistent with the measured value of
$b$(D~I) = 14.0 $\pm$ 1.0 km s$^{-1}$.


We use the column densities of the metal ions 
to deduce the metallicity and neutral fraction in the absorbing gas.
Following Donahue \& Shull (\cite{don91}), assuming a
typical QSO photoionizing spectrum 
given by Mathews \& Ferland (\cite{mat87}),
we estimate the ionization parameter (the ratio of the number of photons 
with energies above one Rydberg to the number of atoms: $n_{\gamma}/n_p$)
is U $\approx 10^{-3.0 \pm 0.5}$ 
in both components.  From the photoionization model, this corresponds to 
a neutral hydrogen fraction of H~I/H $\approx 10^{-2.3 \pm 0.5}$.  
The abundances of the two components, given in the Table, are both low
and they differ.  This is not a surprise because
at early times, and in the outer regions of galaxies, there would be
a large dispersion in abundances in gas which had not mixed.
In both cases [C/Si] $\simeq -0.3$, as is seen
in Galactic stars which have similar low metal abundances.

The low metallicity of this absorption system 
and standard models of chemical evolution 
(\cite{edm94},\cite{ste95},\cite{ste92})
imply negligible destruction of deuterium from its primordial value, since
stars which destroy D and eject H (without D) also eject metals such as Si.


We believe that our measured value of
D/H = 2.3 $ \pm 0.3 \pm 0.3 \times 10^{-5}$
is a detection, and not an upper limit, for the following reasons.
First, the blue side of the \Lya line of the blue component is
very steep, giving a $b$(D) which is lower than 99\% of H lines. 
Even if D/H in the two components is at the ISM value
and additional H at some near by velocity
accounts for most of this absorption, we find this contaminating
H would have $b<18$ \kms, which is low enough to be unusual.
Second, this $b$(D) value agrees with the value predicted from the C~II, Si~II
and H~I lines. Third, our data on this absorption system are sensitive to
D/H $\rm \simeq few \times 10^{-6}$, because $N$(H~I) is large, 
the $b$(H) is low, we see up to Lyman 19 and 8 metal lines (which provide 
the redshifts, the component structure, $b_{therm}$ and $b_{tur}$),
the velocity structure is unusually simple and the signal to noise is high.
Fourth, we have simulated the expected overestimation of D/H due
to contaminating H~I, and find that the expected D/H is virtually identical
to the D/H directly measured. 
This last point applies only to this data, and to this QAS, which is
much better constrained than all other QAS, including those with
possible high D/H.

Data on other absorption systems (\cite{son94},\cite{car94},\cite{wam96},
\cite{car96})
are about 10 times less sensitive because of lower H~I columns 
or lower quality spectra,
and probably would not show
D/H $\ll  10^{-4}$. There exist spectra of
50 -- 100 QSOs which have the sensitivity to detect 
D/H $\simeq 10^{-4}$, and perhaps one in 10 QSOs has an absorption system 
which appears suitable,
so we expect a few false coincidences where an H line lies at the 
expected position of D.
These cases will give high upper limits on D/H because 
the N(H) values are too low to give the sensitivity needed to see low D/H.
While the spectra of Q0014+813 (\cite{son94},\cite{car94}) 
are compatible with our
D/H if their D line is H, our spectrum is completely incompatible with
a high D/H value.

In the local interstellar medium D/H $= 1.6 \pm 0.1 \times 10^{-5}$ 
(\cite{lin95}, \cite{mcc92}). 
If our measured value is the primordial value,
then about 30\% of D has been destroyed, and
about 30\% of atoms in the ISM have been inside stars, consistent with
conventional models of Galactic chemical evolution (\cite{gal95}).
Edmunds (\cite{edm94}) has shown that the maximum depletion of D which 
can occur for arbitrary inflow and outflow of gas from the interstellar 
medium is $X_D/X_{D0} = f^{-1+1/\alpha }$, 
where $X_{D0}$ is the primordial mass fraction of D,
$X_{D}$ is the current mass fraction of D,
$f$ is gas mass divided by the total mass in gas plus stars, about
0.1 -- 0.2 in the solar neighborhood, and
$\alpha $ is the fraction of mass formed into stars which remains in
stellar remnants, conventionally taken as about 0.7 -- 0.8.
We then expect $X_D/X_{D0} \simeq 0.5 - 0.7$, in excellent agreement with our
measurement of about 0.7.
The best estimate of D/H in the gas from which the sun formed is 
($2.7 \pm 1.0$ random $\pm$ 0.5 systematic) $\times 10^{-5}$ (\cite{tur96}).
The errors allow this value to lie between the current ISM value and our 
measurement, as it should.

Our D/H measurement implies a cosmological baryon to photon ratio of
$\rm Log \; \eta = -9.18 \pm 0.4 \pm 0.4 \pm 0.2$ (where the third error is the
1$\sigma$ theoretical uncertainty from BBN predictions, \cite{smi93}), and a 
current density of baryons in terms of the critical density of
$\Omega_b = 0.024 ^{+0.006}_{-0.005} \, h^{-2}$, where
$\rm H_0 = 100\it h \rm ~km$ s$^{-1}$ Mpc$^{-1}$.
Since the observed density of baryons in stars and hot gas is about
0.003 (\cite{per92}), most baryons (about 94\% for $h=0.7$) are 
dark matter, perhaps MACHOS in galaxy halos, and
the hot intergalactic gas which has now been seen but not quantified.
 
Our D/H measurement implies a $^4$He primordial
mass fraction of $Y_p = 0.249 \pm 0.001 \pm 0.001 \pm 0.001$ 
(\cite{kra95},\cite{sar96}), 
which is higher than recent values from extragalactic H~II 
regions; $Y_p = 0.228 \pm 0.005$ (\cite{pag92}) 
and $Y_p = 0.241 \pm 0.003$ (\cite{thu96}).
But it is now realized 
that systematic uncertainties might allow $Y_p =0.25$
(\cite{sas95},\cite{pag94}).
Our measurement also implies Log($^7$Li/H) $= -9.3 \pm 0.1 \pm 0.1 \pm 0.15$,
which is 
between population I and II values, and is allowed
because $^7$Li (\cite{rya96}), like $^3$He (\cite{gal95}), 
is created and destroyed in uncertain amounts.

\acknowledgments
We are extremely grateful to W.M. Keck foundation which made this
work possible, and Tom Bida, Teresa Chelminiak and Wayne Wack for 
assistance at the telescope. 
We thank Bob Carswell and Tom Barlow for superb software, and
Christian Cardall, George Fuller and David Kirkman for many 
useful conversations. 

\clearpage


\begin{table*}
\begin{center}
\begin{tabular}{cccccc}
Ion & Blue Component & Red Component & Total \cr
\tableline
H I & N = 17.74 $\pm$ 0.09 & N = 17.50 $\pm$ 0.19 & N = 17.94 $\pm$ 0.05 \cr
    & b = 17.1 $\pm$ 0.7 & b = 22.2 $\pm$ 3.0 & \cr
D I & N = 13.10 $\pm$ 0.08 & N = 12.86 (tied) & N = 13.30 $\pm$ 0.04 \cr
    & b = 14.0 $\pm$ 1.0 & b = 16.7 (tied) & \cr
Si II & N = 11.76 $\pm$ 0.07 & N = 12.41 $\pm$ 0.02 & N = 12.50 $\pm$ 0.02 \cr
    & b = 5.4 $\pm$ 1.6 & b = 9.6 $\pm$ 0.8 & \cr
Si III & N = 12.73 $\pm$ 0.20 & N = 13.20 $\pm$ 0.05 & N = 13.33 $\pm$ 0.04 \cr
    & b = 5.5 $\pm$ 1.6 & b = 14.3 $\pm$ 2.0 & \cr
Si IV & N = 12.02 $\pm$ 0.12 & N = 13.03 $\pm$ 0.02 & N = 13.07 $\pm$ 0.02 \cr
    & b = 3.8 $\pm$ 1.7 & b = 10.6 $\pm$ 0.5 & \cr
C II & N = 12.70 $\pm$ 0.08 & N = 13.27 $\pm$ 0.03 & N = 13.37 $\pm$ 0.02 \cr
    & b = 7.0 $\pm$ 1.2 & b = 10.0 $\pm$ 0.5 & \cr
C III & N = 13.36 $\pm$ 0.09 & N = 13.82 $\pm$ 0.19 & N = 13.95 $\pm$ 0.15 \cr
    & b = 7.9 $\pm$ 1.6 & b = 12.3 $\pm$ 1.1 & \cr
C IV & N = 12.22 $\pm$ 0.16 & N = 12.61 $\pm$ 0.11 & N = 12.76 $\pm$ 0.08 \cr
    & b = 13.1 $\pm$ 5.9 & b = 9.2 $\pm$ 2.2 & \cr
O I & ... & ... & N $< 12.6$ (2$\, \sigma$) \cr
T($10^4$ K) & 1.62 $\pm$ 0.09 & 2.36 $\pm$ 0.09 & \cr
b$_{tur}$ & 4.8 $\pm$ 0.8 & 8.4 $\pm$ 0.4 & \cr
[C/H]     & $-3.0\pm 0.3$ & $-2.2 \pm 0.3$ & ... \cr
[Si/H]     & $-2.8\pm 0.2$ & $-1.9 \pm 0.2$ & ... \cr
[C/Si]     & $-0.2$        & $-0.3$ & ... \cr
\end{tabular}
\end{center}
 
\tablenum{1}
\caption{Column Densities and Velocity Dispersions: Q1937-1009, $z=3.572$}
 
\end{table*}

\clearpage

\section{Table Captions}

Table 1:  Results from the simultaneous fit of the two component model
of the absorption system at $z$ = 3.572 towards QSO 1937--1009.
Column densities are in (cm$^{-2}$) and $b$ values in (km s$^{-1}$).
The 1$\sigma$ random errors represent deviations which change 
the $\chi^2$ of fits by 1$\sigma$.
The absorption line parameters were taken from Morton (\cite{mor91}).
The last column shows the sum of the column densities in both components
which are much better determined
than the distribution of the gas between the components.
Hence the errors on the sums are smaller than on the individual components,
and the best determination of D/H is the total D~I column
divided by the total H~I column in the two components.

The abundances of C and Si are presented in logarithmic units
relative to solar, where e.g. [C/H] $= -3$, means $10^{-3}$ of the solar C
abundance.
The upper limit on O~I (1302\AA) gives [O/H] $< -1.5$, for $U=10^{-3}$.
One $b$ value for each element at each velocity component 
provides a good fit to 
all the metal lines ($\chi^2 = 10.6$ for 8 degrees of freedom), and especially 
to the important blue components. Although the red components of the 
saturated  Si~III and C~III lines both have $b$ values a bit larger than 
the mean, this is probably not significant because we do not see a trend 
with ionization.

\section{Figure Captions}
 
Figure 1: Velocity plots of Lyman series lines (left), and
all the metal lines
(right) in the absorption system towards QSO 1937--1009. 
Zero velocity corresponds to the redshift $z = 3.572201$ of the blue component. 
The red component at $z = 3.572428$ is indicated by a second dashed line at
$+15$ km s$^{-1}$.  The deuterium lines lie 82 \kms to the blue of their
respective hydrogen lines. 
The histogram represents the observed counts of the combined Keck spectra
in each pixel, normalized to the
quasar continuum.  The smooth curve shows the Voigt profiles convolved
with the instrumental resolution which produces the best 
simultaneous fit to all the lines.  The values and $1\sigma$ errors which 
characterize the smooth curve are
summarized in Table 1.

Levshakov and others (\cite{lev96},\cite{gai74})
have developed the mesoturbulent theory which gives absorption 
line profiles which differ from normal Voigt profiles
when $b_{tub} \geq b_{therm}$ and the turbulent velocity
fields are correlated on distances less the cloud size.
In the limit when thermal velocities (which are necessarily uncorrelated)
dominate, mesoturbulence predicts the Voigt profiles which we have
used.  In all hydrogen and deuterium lines of our system
$b_{tub} < 0.4 b_{therm}$, so we conclude that Voigt profiles
are adequate.

The \Ly series lines, C~III (977), Si~II (1260) and
Si~III (1206) are all in the Lyman alpha forest region where there is 
abundant absorption, usually from cosmologically distant and unrelated H~I. 
We do not fit this additional absorption, which is seen in
Si~III (1206) and D \Lyb, and we omit the most blended \Ly lines.

The signal to noise per pixel ($\rm \approx 4 \, km \, s^{-1}$) 
for the Lyman series from top to bottom is 
75, 31, 22, 19, 17, 15, 15 and 13.
For the metal lines, from top to bottom, it is
100, 20, 71, 40, 56, 23, 50 and 50.

Figure 2: 
Lyman limit portion of the spectrum of QSO 1937--1009. The
histogram represents measured flux in each pixel, 
normalized to the continuum level, and wavelength calibrated 
to heliocentric vacuum values. 
The smooth curve shows the Voigt profiles characterized by the
values in Table 1.

The spectra were obtained in 5 exposures covering 3890 -- 7450 \AA\ on
July 31 and Aug. 1, 1994.
We used a Textronix 2303x2048 CCD with 2x2 pixel on chip binning.
A 1.14 arcsec slit was used to give a resolution of about 
8 kms$^{-1}$ measured from unblended lines in calibration spectra. 
All exposures covered 5045 -- 6300 \AA\ , which includes \Lya, and we added
overlapping spectra, weighting each pixel by the square of its signal-to-noise.
Each exposure of QSO 1937--1009 was accompanied by
dark, quartz lamp, and Thorium-Argon arc-lamp exposures.  A standard star
was also observed before or after exposures to trace the echelle
orders and remove the blaze response in the spectra.
The one-dimensional spectra were extracted from the two-dimensional images
using an automated echelle extraction routine which
performs baseline subtraction,
bias subtraction, and flat-field correction before extracting the orders.
The program also masks regions of the CCD subject to defects
(e.g. the ink spot, hot columns) and pixels struck by cosmic rays, and it
automatically finds wavelengths.

The continuum level redward of Lyman-alpha emission
came from a sixth order cubic spline with 3 sigma high
and low pixel rejection.  Between
Lyman-alpha emission (5820 \AA) and the Lyman continuum break (4170 \AA),
we used a third, fourth, or fifth order Legendre polynomial,
depending on the severity of the Lyman-alpha forest absorption, and we
rejected pixels outside the range $-1.2$ to $+3\sigma$.

\begin{figure}
\figurenum{1}
\centerline{
\psfig{figure=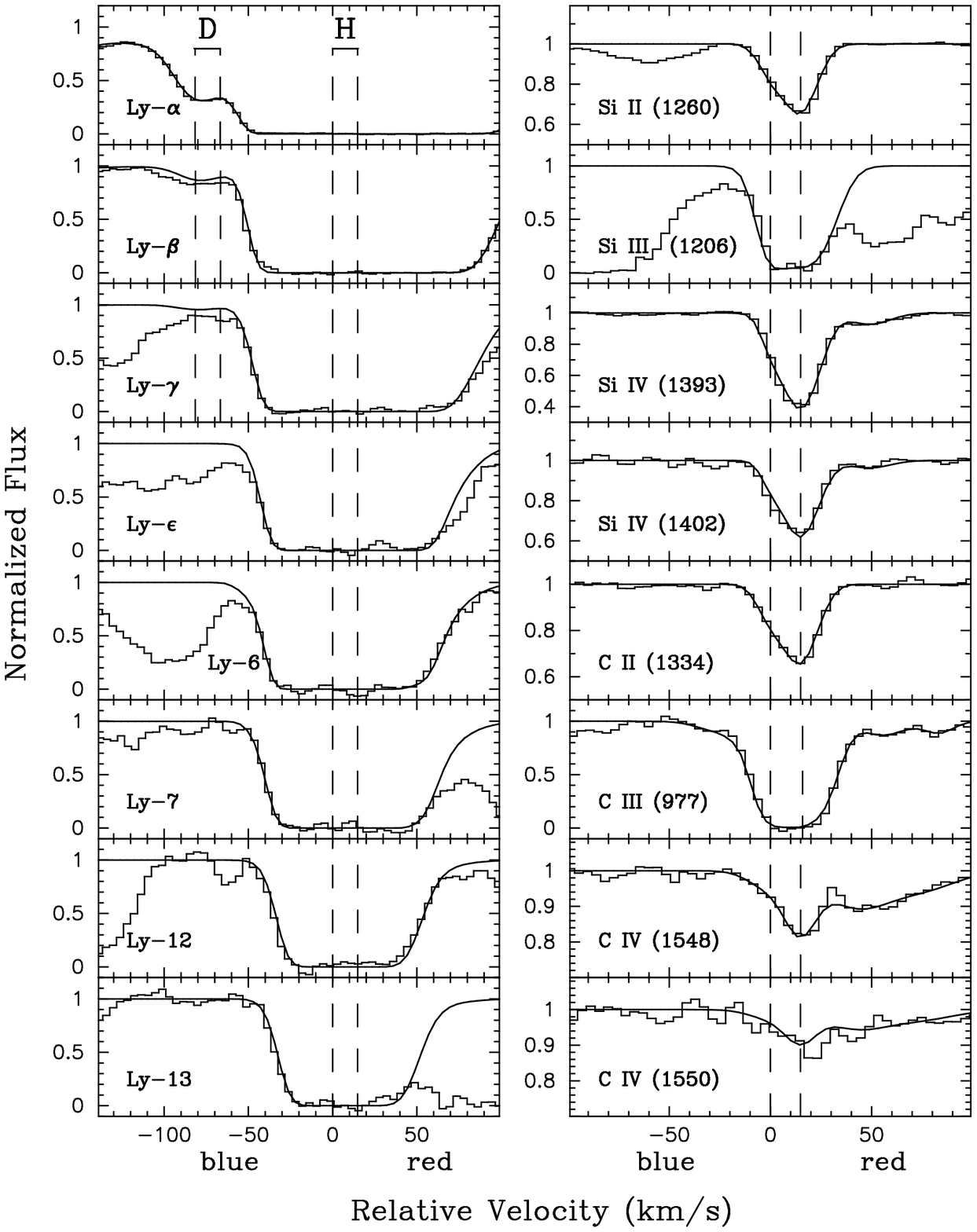,height=9.5in}}
\end{figure}

\begin{figure}
\figurenum{2}
\centerline{
\psfig{figure=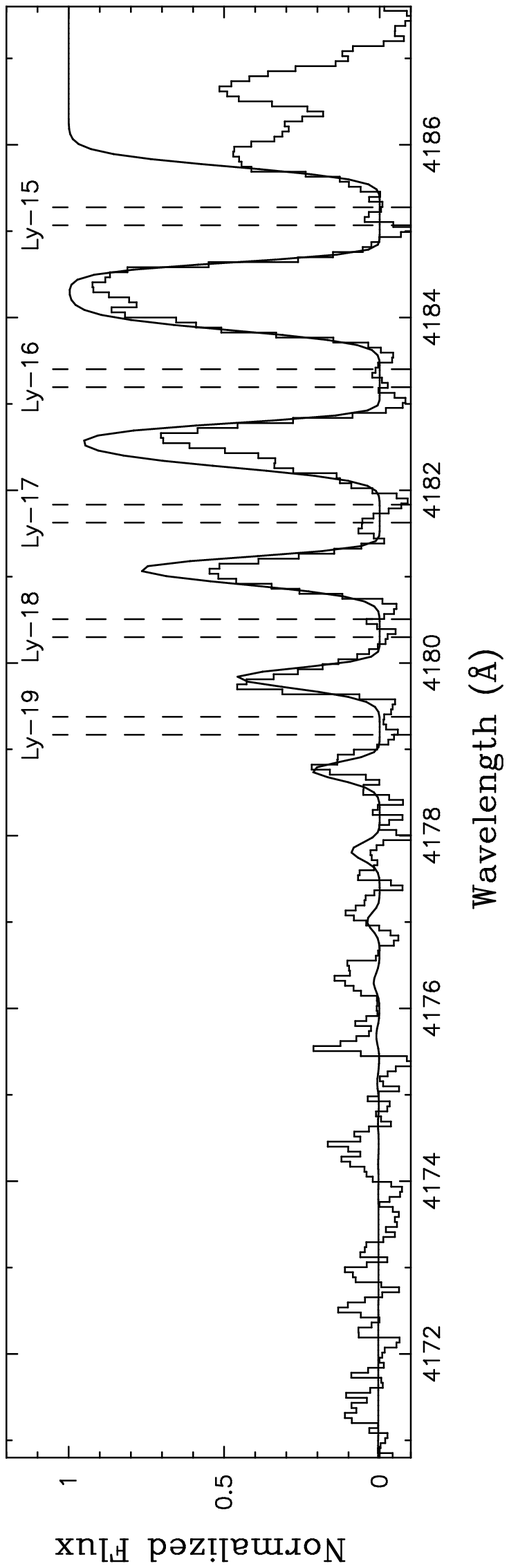,height=9.5in}}
\end{figure}

\end{document}